\documentclass[a4paper,11pt]{article}
\usepackage{pos}

\def\onehalf{\hbox{$\frac{1}{2}$}}
\def\go{\rightarrow}
\def\dd{\partial}

\def\KK{{\rm KK}}

\def\la{\langle}
\def\ra{\rangle}
\def\flat{{\rm flat}}
\def\RS{{\rm RS}}
\def\Tr{{\rm Tr} \,}
\def\Psibar{\overline{\Psi}}

\makeatletter
\@addtoreset{equation}{section}
\makeatother

\title{Holography in anomaly flow in orbifold gauge theory}

\author*{Yutaka Hosotani}

\affiliation{Research Center for Nuclear Physics, Osaka University\\
Ibaraki, Osaka 567-0047, Japan}
  
\emailAdd{hosotani@rcnp.osaka-u.ac.jp}

\abstract{
In orbifold gauge theory and gauge-Higgs unification models, gauge anomaly flows with an Aharonov-Bohm 
phase $\theta_H$ in the fifth dimension.
We  analyze $SU(2)$ gauge theory with doublet fermions  in the flat  $M^4 \times (S^1/Z_2)$  spacetime 
and in the Randall-Sundrum (RS) warped  space. 
With orbifold boundary conditions the $U(1)$ part of gauge symmetry remains unbroken 
at $\theta_H = 0$ and $\pi$.
Chiral anomalies smoothly vary with $\theta_H$ in the RS space.
Anomaly coefficients associated with this anomaly flow are expressed
in terms of the values of the wave functions of gauge fields at the  UV and IR branes in the RS space
and parity conditions of fermion fields.
Holography in anomaly flow is observed.  
Conditions for the anomaly cancellation turn out independent of $\theta_H$.
}

\FullConference{%
  Corfu Summer Institute 2022 "School and Workshops on Elementary Particle Physics and Gravity",\\
  28 August - 1 October, 2022\\
  Corfu, Greece}

\begin{document}
\maketitle

\section{Introduction}

Chiral fermions generally induce chiral anomaly.\cite{Adler1969, BellJackiw1969, Fujikawa1979}
 Even for massive fermions
chiral anomaly can be generated if gauge couplings are not purely vector-like. 
Something special happens in gauge-Higgs unification (GHU), in which gauge symmetry is dynamically broken
by an Aharonov-Bohm (AB) phase $\theta_H$ in the fifth dimension.\cite{Hosotani1983, Davies1988, Hatanaka1998}
It has been noticed in the GUT inspired $SO(5) \times U(1)_X \times SU(3)_C$ GHU in the Randall-Sundrum (RS) 
warped space \cite{FiniteT2021} that quarks and leptons are massless and purely chiral at $\theta_H=0$, 
become massive at $\theta_H \not= 0$,  and smoothly become vector-like at $\theta_H=\pi$.  
Gauge bosons such as $W$ and $Z$ bosons are massless gauge bosons of 
$SU(2)_L \times U(1)_Y$ at  $\theta_H=0$, become massive at $\theta_H \not= 0$, and smoothly converted
to massless gauge bosons of $SU(2)_R \times U(1)_{Y'}$ at  $\theta_H=\pi$.
This prompts the following question.  What happens to the anomaly generated by chiral quarks and leptons at 
$\theta_H=0$?  Does it disappear at $\theta_H=\pi$?  What is the fate of chiral anomaly?
 
To pin down what is going on in GHU formulated on orbifolds, the dependence of chiral  anomaly on the
AB phase $\theta_H$ in $SU(2)$ gauge theory with doublet fermions
in the flat $M^4 \times (S^1/Z_2)$ space and in the RS warped space 
has been investigated in Refs.\ \cite{AnomalyFlow1, AnomalyFlow2}.  In the flat  $M^4 \times (S^1/Z_2)$ space 
the mass spectrum of the Kaluza-Klein (KK) modes of gauge bosons and fermions changes
linearly in $\theta_H$, namely as $m_n = | n + (\theta_H/\pi)|/R$ or $ | n + (\theta_H/2\pi)|/R$,  so that the level crossing 
takes place where $\theta_H $ is a multiple of $\onehalf \pi$ or $\pi$.
In the RS space there occurs no level crossing in the mass spectrum.  The spectrum varies smoothly as $\theta_H$ 
changes from 0 to $2\pi$.  The lowest mode remains as the lowest mode for any $\theta_H$.
Gauge couplings of right- and left-handed modes of fermions vary smoothly as $\theta_H$,
and the magnitude of chiral anomaly also changes as $\theta_H$.  The anomaly coefficient (defined below) 
coming from  one doublet fermion changes from $2$ at $\theta_H=0$ to $0$ at $\theta_H=\pi$.
The anomaly flows with the AB phase $\theta_H$.

Furthermore it was shown that  the magnitude of the total anomaly evaluated by summing  
contributions coming from all KK modes of fermions running along  internal loops is expressed 
in terms of the values of the wave functions of the gauge fields at the two branes (the UV and IR branes 
in the RS space) and the parity conditions of fermion fields at the two branes.
Each fermion field in the RS space is characterized by its own bulk mass parameter $c$ which controls
its mass and wave function at general $\theta_H$.   Although the anomaly coming from each KK mode 
depends on the bulk mass parameter $c$, the total anomaly does not depend on $c$.
There emerges a holographic formula for the total anomaly.
This holography becomes crucial to have the cancellation of gauge anomalies in GHU.

\section{$SU(2)$ GHU in $M^4 \times (S^1/Z_2)$}

Let us first consider $SU(2)$ GHU in the flat $M^4 \times (S^1/Z_2)$ spacetime with coordinate 
$x^M$ ($M=0,1,2,3,5$, $x^5 =y$) whose action is given by
\begin{align}
I_\flat &= \int d^4 x \int_0^{L} dy  \,  {\cal L}_\flat ~, ~~
{\cal L}_\flat  =
- \frac{1}{2} \Tr  F_{MN} F^{MN} 
+  \Psibar \gamma^M ( \dd_M - i g_A A_M ) \Psi    ~,
\label{flataction}
\end{align}
where ${\cal L}_\flat (x^\mu, y)= {\cal L}_\flat (x^\mu, y + 2L) = {\cal L}_\flat (x^\mu, -y) $.
$A_M = \onehalf  \sum_{a=1}^3 A^a_M \tau^a$ and $F_{MN} = \dd_M A_N - \dd_N A_N - i g_A [A_M , A_N]$
where $\tau^a$'s are Pauli matrices. 
Orbifold boundary conditions are given, with $(y_0, y_1) = (0, L)$ and $P_0 = P_1 = \tau^3$, by
\begin{align}
&\begin{pmatrix} A_\mu \cr A_y \end{pmatrix} (x, y_j - y) 
= P_j \begin{pmatrix} A_\mu \cr - A_y \end{pmatrix} (x,  y_j + y) P_j^{-1} ~, \cr
\noalign{\kern 5pt}
&\Psi  (x, y_j - y)  = 
\begin{cases} 
+ P_j \gamma^5 \Psi  (x, y_j + y) &\hbox{type 1A} \cr
\noalign{\kern 1pt}
-  P_j \gamma^5 \Psi  (x, y_j + y) &\hbox{type 1B} \cr
\noalign{\kern 1pt}
(-1)^j P_j  \gamma^5  \Psi  (x, y_j + y) &\hbox{type 2A} \cr
\noalign{\kern 1pt}
(-1)^{j+1} P_j  \gamma^5  \Psi  (x, y_j + y) &\hbox{type 2B} 
 \end{cases}  ~.
\label{BC1}
\end{align}
The $SU(2)$ symmetry is broken to $U(1)$.  
$A^3_\mu$ and $A^{1,2}_y$ are parity even at both $y_0$ and $y_1$.
The zero mode of $A^3_\mu$ is the 4D $U(1)$ gauge field.
The 4D gauge coupling is given by $g_4 = g_A/\sqrt{L}$.
The zero modes of  $A_y^{1,2}$ may develop  nonvanishing expectation values.
Without loss of generality we assume $\la A_y^{1} \ra = 0$. 
An AB phase $\theta_H$ along the fifth dimension is then given by
\begin{align}
&P \exp \bigg\{  i g_A \int_0^{2 L} dy \, \la A_y \ra \bigg\} = e^{ i \theta_H \tau^2 } ~,~~
\theta_H = g_4 L \, \la  A_y^{2} \ra ~.
\label{ABphase1}
\end{align}

When $\theta_H \not= 0$, $A_\mu^1$ and $A_\mu^3$ intertwine with each other.  
It is straightforward to find mass eigenstates.  The KK expansion is given by
\begin{align}
\begin{pmatrix} A_{\mu}^{1} (x,y) \cr \noalign{\kern 2pt}  A_{\mu}^{3} (x,y) \end{pmatrix} 
&= \sum_{n=-\infty}^{\infty } B_{\mu}^{(n )}  (x) 
\frac{1}{\sqrt{\pi R}}
\begin{pmatrix} \sin (ny/R) \cr \cos  (ny/R) \end{pmatrix} 
\label{gaugeKKflat1}
\end{align}
where $R = L/\pi$.  The mass of the $B_{\mu}^{(n)}  (x)$ mode is 
$m_{n} (\theta_{H}) =R^{-1}\big| n + \frac{\theta_{H}}{\pi} \big|$.  
The spectrum is periodic in $\theta_H$ with a period $\pi$.
The KK expansion of a doublet fermion $\Psi=(u,d)^t$ of type 1A is given by
\begin{align}
\begin{pmatrix} u_{R} (x,y) \cr \noalign{\kern 2pt} d_{R} (x,y) \end{pmatrix} 
&= \sum_{n=-\infty}^{\infty } \psi_{R}^{(n)}  (x) 
\frac{1}{\sqrt{\pi R}}
\begin{pmatrix} \cos (ny/R)   \cr 
 \sin(ny/R)  \end{pmatrix}, \cr
\noalign{\kern 5pt}
\begin{pmatrix}u_{L} (x,y) \cr \noalign{\kern 2pt} d_{L} (x,y) \end{pmatrix} 
&= \sum_{n=-\infty}^{\infty } \psi_{L}^{(n)}  (x) 
\frac{1}{\sqrt{\pi R}}
\begin{pmatrix} - \sin (ny/R)  \cr 
 \cos (ny/R)   \end{pmatrix}.
\label{fermionKKflat1}
\end{align}
$\psi_{R}^{(n)}  $ and $\psi_{L}^{(n)} $ combine to form
the $\psi^{(n)}  (x) $ mode with a mass  given by 
$m_{n}(\theta_{H}) = R^{-1} \big| n + \frac{\theta_{H}}{2 \pi} \big|$. 
The spectrum is periodic in $\theta_H$ with a period $2\pi$.
Note that the KK mass scale in the flat space is $m_\KK^\flat = 1/R$.

\section{$SU(2)$ GHU in the Randall-Sundrum warped space}

The metric of the RS warped space is given by \cite{RS1999}
\begin{align}
ds^2= e^{-2\sigma(y)} \eta_{\mu\nu}dx^\mu dx^\nu+dy^2
\label{RSmetric1}
\end{align}
where $\eta_{\mu\nu}=\mbox{diag}(-1,+1,+1,+1)$, $\sigma(y)=\sigma(y+ 2L)=\sigma(-y)$ 
and $\sigma(y)=ky$ for $0 \le y \le L$.  It has the same topology as $M^4 \times (S^{1}/Z_{2})$.
In the region $0 \le y \le L$ the metric can be written, in terms of the conformal coordinate 
$z = e^{ky}$, as
\begin{align}
ds^2=  \frac{1}{z^2} \bigg(\eta_{\mu\nu}dx^{\mu} dx^{\nu} + \frac{dz^2}{k^2}\bigg) 
\quad ( 1 \le z \le z_L= e^{kL}) ~.
\label{RSmetric2}
\end{align}
$z_L$ is called the warp factor of the RS space.
The RS space is an anti-de Sitter (AdS) space sandwiched by the UV brane at $y=0 ~(z=1)$ 
and the IR brane at $y=L~(z=z_L)$.   The AdS curvature is given by $\Lambda = - 6 k^2$.

The action in the RS space is given by
\begin{align}
&I_\RS = \int d^5 x  \sqrt{- \det G}   \, {\cal L}_\RS ~, ~~
{\cal L}_\RS  = - \frac{1}{2} \Tr  F_{MN} F^{MN}  +  \Psibar {\cal D} (c) \Psi   ~, \cr
\noalign{\kern 3pt}
&{\cal D} (c) =  \gamma^A {e_A}^M
\Big( \dd_M - i g_AA_M +\frac{1}{8}\omega_{MBC}[\gamma^B,\gamma^C]  \Big) - c \, \sigma' ~.
\label{RSaction}
\end{align}
$c$ is a dimensionless bulk mass parameter.
Note ${\cal L}_\RS (x^\mu, y) = {\cal L}_\RS  (x^\mu, -y) = {\cal L}_\RS (x^\mu, y + 2L)$.
Fields $A_{M}$ and $\Psi$ satisfy the same boundary conditions (\ref{BC1}) as in the flat spacetime.
The AB phase $\theta_H$ and  the  KK mass scale $m_\KK$ are given by 
\begin{align}
\theta_{H} &= \frac{\la A_{z}^{2 (0)} \ra}{f_{H}} ~, ~~
f_{H} = \frac{1}{g_{4}} \sqrt{ \frac{2k}{L(z_{L}^{2} - 1)} } ~, ~~
m_\KK = \frac{\pi k}{z_L-1} ~,
\label{ABphase2}
\end{align}
where $A_{z}^{a} (x,z) = k^{-1/2} \sum A_{z}^{a (n)} (x) v_{n}(z)$ and $v_{0} (z) = \sqrt{2/(z_{L}^{2}-1)} \,  z$.

The KK expansion of the gauge fields $A_\mu^1$ and 
$A_\mu^3$ is given by 
\begin{align}
&\begin{pmatrix} A_{\mu}^{1} (x,y) \cr \noalign{\kern 1pt}  A_{\mu}^{3} (x,y) \end{pmatrix} 
=  \frac{1}{\sqrt{L}} \sum_{n=0}^{\infty } Z_{\mu}^{(n)} (x) \,  \begin{pmatrix}  h_n(y) \cr  k_n (y) \end{pmatrix}, \cr
\noalign{\kern 5pt}
&\begin{pmatrix}  h_n(y) \cr  k_n (y) \end{pmatrix} 
= \begin{pmatrix}  - h_n(- y) \cr  k_n (- y) \end{pmatrix}
= \begin{pmatrix}  h_n(y+ 2L) \cr  k_n (y + 2L) \end{pmatrix} \cr 
\noalign{\kern 3pt}
&
= \begin{pmatrix} \cos \theta (z) & \sin \theta (z) \cr - \sin \theta (z) & \cos \theta (z) \end{pmatrix} 
\begin{pmatrix} \tilde h_n(z) \cr \tilde k_n (z) \end{pmatrix}, 
~~\theta (z) =   \theta_{H} \, \frac{z_{L}^{2} - z^{2}}{z_{L}^{2} - 1} 
 \quad {\rm for~} 0 \le y \le L ~.
\label{RSgaugeKK1}
\end{align}
The mass spectrum $\{ m_n = k \lambda_n ; ~  \lambda_0 <  \lambda_1 <  \lambda_2< \cdots \}$ of 
the KK modes $\big\{  Z_{\mu}^{(n)} (x) \big\}$ is determined by the zeros of
\begin{align}
Z_\mu^{(n)} : ~ S C' (1; \lambda_{n}) + \lambda_{n} \sin^{2} \theta_{H} = 0 
\label{RSgaugespectrum1}
\end{align}
where $S (z; \lambda)$ and  $C (z; \lambda)$ are expressed in terms of Bessel functions and 
are given by (\ref{functionA1}).
The wave functions $\tilde {\bf h}_n(z) \equiv ( \tilde h_n(z) ,  \tilde k_n(z) )^t$ are given by (\ref{RSgaugeKKB1}).

For fermion fields we define $\check \Psi (x,z)= z^{-2} \,  \Psi  (x,z)$ for $1 \le z \le z_L$.  
The KK expansion of $\check \Psi = (\check u, \check d)^t$ of type 1A  is given by
\begin{align}
\begin{pmatrix}{\check u}_{R} (x,y) \cr \noalign{\kern 2pt} {\check d}_{R} (x,y) \end{pmatrix} 
&= \sqrt{k} \sum_{n= 0}^\infty   \chi_{R}^{(n)} (x)    \begin{pmatrix}  f_{Rn} (y) \cr   g_{Rn} (y) \end{pmatrix},  \cr
\noalign{\kern 5pt}
\begin{pmatrix}  {\check u}_{L} (x,y) \cr \noalign{\kern 2pt} {\check d}_{L} (x,y) \end{pmatrix} 
&= \sqrt{k} \sum_{n= 0}^\infty   \chi_{L}^{(n)} (x)  \begin{pmatrix}   f_{Ln} (y) \cr   g_{Ln} (y) \end{pmatrix},
\label{RSfermionKK1}
\end{align}
where
\begin{align}
&\begin{pmatrix}  f_{Rn} (y) \cr   g_{Rn} (y) \end{pmatrix}
=  \begin{pmatrix}  f_{Rn} (- y) \cr   - g_{Rn} (- y) \end{pmatrix} 
=  \begin{pmatrix}  f_{Rn} (y+ 2L) \cr   g_{Rn} (y+ 2L) \end{pmatrix}  \cr
\noalign{\kern 3pt}
&\quad
=  \begin{pmatrix} \cos \onehalf \theta (z) & - \sin \onehalf \theta (z) \cr 
\noalign{\kern 1pt}
 \sin\onehalf  \theta (z) & \cos \onehalf \theta (z) \end{pmatrix} 
 \begin{pmatrix}  \tilde f_{Rn} (z) \cr   \tilde g_{Rn} (z) \end{pmatrix}  \quad {\rm for~} 0 \le y \le L ~, \cr
\noalign{\kern 5pt}
&\begin{pmatrix}  f_{Ln} (y) \cr   g_{Ln} (y) \end{pmatrix}
=  \begin{pmatrix}  - f_{Ln} (- y) \cr    g_{Ln} (- y) \end{pmatrix} 
=  \begin{pmatrix}  f_{Ln} (y+ 2L) \cr   g_{Ln} (y+ 2L) \end{pmatrix}  \cr
\noalign{\kern 3pt}
&\quad 
=  \begin{pmatrix} \cos \onehalf \theta (z) & - \sin \onehalf \theta (z) \cr 
\noalign{\kern 1pt}
 \sin\onehalf  \theta (z) & \cos \onehalf \theta (z) \end{pmatrix} 
 \begin{pmatrix}  \tilde f_{Ln} (z) \cr   \tilde g_{Ln} (z) \end{pmatrix}  \quad {\rm for~} 0 \le y \le L ~.
\label{RSfermionKK2}
\end{align}
$\chi_{R}^{(n)}  (x)$ and $\chi_{L}^{(n)} (x)$ combine to form a massive mode $\chi^{(n)}  (x)$ for $\theta_H \not= 0$.
The mass spectrum $\{ m_n = k \lambda_n ; ~  \lambda_0 <  \lambda_1 <  \lambda_2< \cdots \}$ of 
the KK modes $\{  \chi^{(n)} (x) \}$ is determined by the zeros of
\begin{align}
 \chi^{(n)} : ~ S_{L} S_{R}(1; \lambda_{n}, c) +  \sin^{2} \onehalf \theta_{H} = 0  
\label{RSfermispectrum1}
\end{align}
where $S_L (z; \lambda,c)$ and  $S_R (z; \lambda, c)$ are given by (\ref{functionA3}).
The wave functions $\tilde{\bf f}_{Rn} (z) = (\tilde f_{Rn} (z), \tilde g_{Rn} (z) )^t$ and
$\tilde{\bf f}_{L n} (z) = (\tilde f_{L n} (z), \tilde g_{L n} (z) )^t$ are given by (\ref{RSfermionKKB1}).

\begin{figure}[bth]
\centering
\includegraphics[height=40mm]{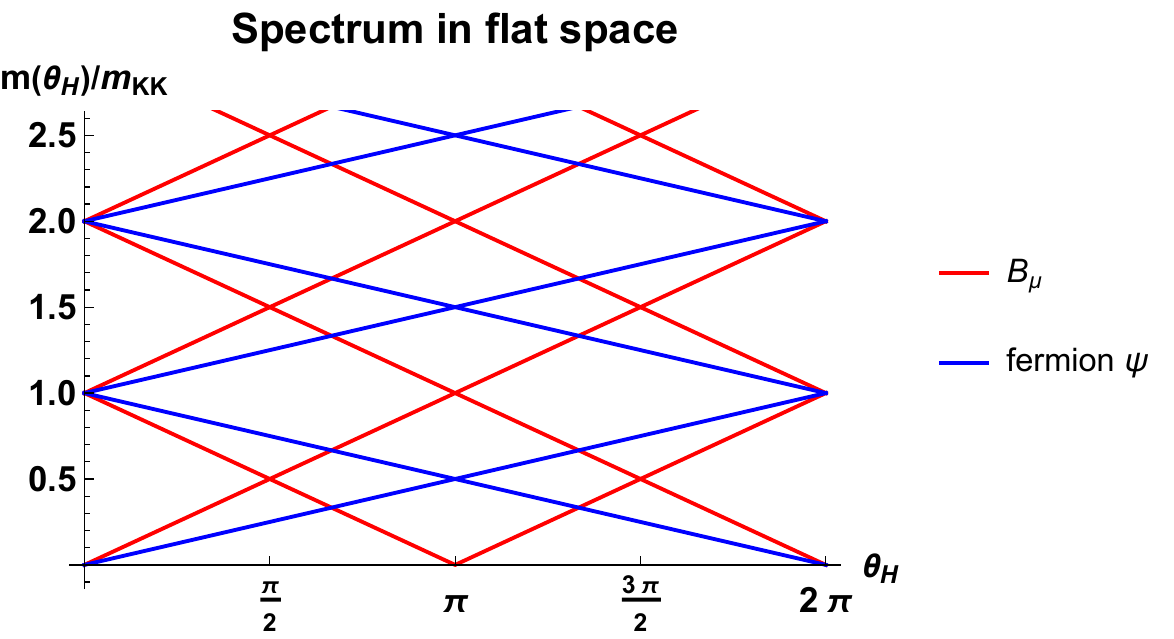}
\vskip 5pt
\includegraphics[height=40mm]{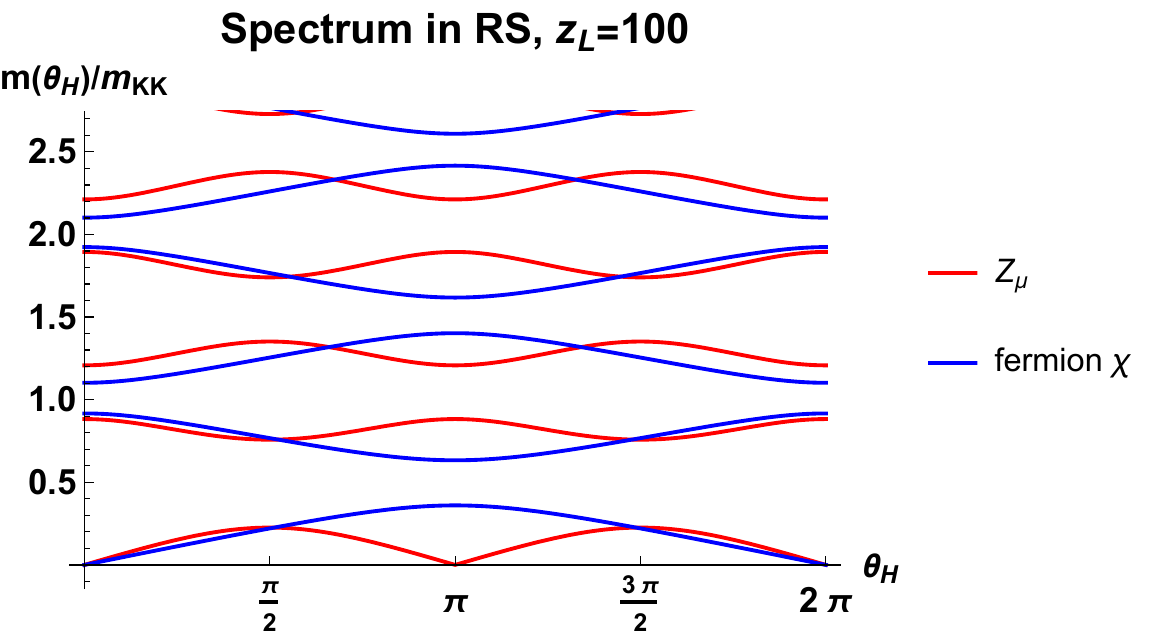}
\caption{(Top):  The mass spectrum of gauge fields $B_{\mu}^{(n)}$ and fermion  fields $\psi^{(n)}$ (type 1A)
in the flat $M^4 \times (S^1/Z_2)$ space is displayed. The level crossing occurs at $\theta_H =0, \onehalf \pi, \pi, \cdots$. 
(Bottom): The mass spectrum of gauge fields $Z_{\mu}^{(n)}$ and fermion  fields $\chi^{(n)}$ (type 1A)
in the RS warped space is displayed.  The warp factor is $z_L = 100$ and the bulk mass parameter 
of $\Psi$ is $c=0.25$.  
}   
\label{fig:massspectrum}
\end{figure}

The mass spectra as functions of $\theta_H$ in the flat $M^4 \times (S^1/Z_2)$ space and in the RS 
warped space are depicted in Fig.\ \ref{fig:massspectrum}.
In the flat  space the mass spectrum of each field  changes linearly in $\theta_H$ so that the level
crossing occurs.  In the RS space there is no level crossing so that physical quantities change smoothly
as $\theta_H$.   It is expected that in the flat space  something singular may occur 
at $\theta_H =0, \onehalf \pi, \pi, \cdots$.   This is exactly what is going to happen in the anomaly as is seen below.

\section{Gauge couplings and anomaly in flat space}

\begin{figure}[bth]
\centering
\includegraphics[height=35mm]{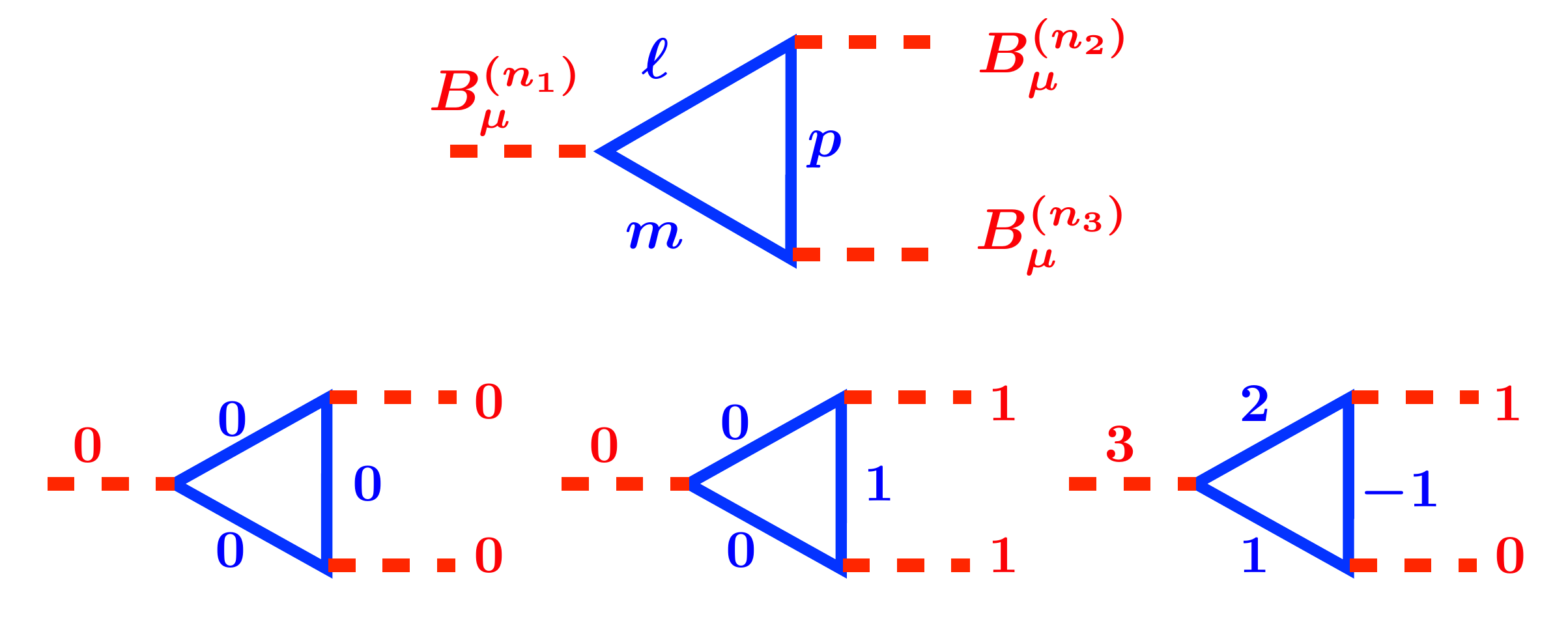}
\caption{Chiral anomaly in the flat space. 
}   
\label{fig:flatanomaly}
\end{figure}

Gauge couplings of fermions in the flat space are easily found by inserting the KK expansions 
(\ref{gaugeKKflat1}) and (\ref{fermionKKflat1}) into the action (\ref{flataction}). 
One finds that the $B_\mu^{(n)}$ couplings of the fermion fields are given by
\begin{align}
& \sum_{n=-\infty}^{\infty} B_{\mu}^{\la n \ra} \, j^\mu_{(n)}  \cr
&=\frac{g_{4}}{2} \sum_{n=-\infty}^{\infty}  \sum_{m=-\infty}^{\infty}   \sum_{\ell=-\infty}^{\infty} 
B_{\mu}^{\la n \ra}  \Big\{   s^{R}_{n m \ell} \,
\psi_{R}^{\la m \ra \, \dagger} \bar \sigma^{\mu} \psi_{R}^{\la \ell \ra}
+  s^{L}_{n m \ell} \, \psi_{L}^{\la m \ra \, \dagger}  \sigma^{\mu} 
\psi_{L}^{\la \ell \ra} \Big\}  ~, \cr
\noalign{\kern 3pt}
&\hskip 3.cm 
s^{R}_{n m \ell} = s^{L}_{n m \ell} = \delta_{n, m+\ell} ~.
\label{flatcoupling1}
\end{align}
Here we have adopted the two-component notation; 
 $\sigma^{\mu}  =( I_{2}, \vec \sigma )$ and  $\bar \sigma^{\mu}  =(- I_{2}, \vec \sigma )$.
 Chiral anomaly in $\dd_\mu  j^\mu_{(n)}$ arises from triangle diagrams  in which various combinations
 of $ \psi_{R/L}^{\la \ell \ra}$ run; 
 \begin{align}
&\dd_\mu  j^\mu_{(n)} + \cdots = -\Big( \frac{g_4}{2}\Big)^3 \sum_\ell  \sum_m 
\frac{b_{n\ell m}}{16\pi^2} \, B_{\mu\nu}^{(\ell)} \tilde B^{(m)\mu\nu}
\label{flatanomaly1}
\end{align}
where $B_{\mu\nu}^{(\ell)} = \dd_\mu B_\nu^{(\ell)}-  \dd_\nu B_\mu^{(\ell)}$.
The anomaly coefficient $b_{n\ell m}$ is found to be
\begin{align}
b_{n_1 n_2 n_3} &= \sum_{m, \ell, p} \Big\{ s_{n_1  m \ell}^R s_{n_2 \ell p}^R  s_{n_3 p m}^R 
+ s_{n_1  m \ell}^L s_{n_2 \ell p}^L  s_{n_3 p m}^L \Big\} \cr
\noalign{\kern 3pt}
&= \begin{cases} 2 &{\rm for~} n_1 + n_2 +n_3={\rm even} \cr
0 &{\rm for~} n_1 + n_2 +n_3={\rm odd} \end{cases} ~.
\label{flatanomaly2}
\end{align}
Chiral anomalies arise even for  $j^\mu_{(n \not= 0)}$.  
A few examples are shown in Fig.\  \ref{fig:flatanomaly}.

\section{Gauge couplings and anomaly in RS}

Similarly gauge couplings of fermions in the RS space are expressed as
\begin{align}
& \sum_{n=0}^{\infty} Z_{\mu}^{\la n \ra} \, j^\mu_{(n)}  \cr
&=\frac{g_4}{2} \sum_{n=0}^\infty \sum_{\ell = 0}^\infty \sum_{m=0}^\infty Z_\mu^{(n)} 
\Big\{ t^R_{n\ell m} \, \chi_R^{(\ell) \dagger} \bar \sigma^\mu \chi_R^{(m)} 
+ t^L_{n\ell m} \, \chi_L^{(\ell) \dagger}  \sigma^\mu \chi_L^{(m)}  \Big\} ~.
\label{RScoupling1}
\end{align}
The couplings $t^R_{n\ell m}$ and $t^L_{n\ell m}$ are more involved.  They are given,
in terms of the wave functions in (\ref{RSgaugeKK1}) and (\ref{RSfermionKK2}),  by
\begin{align}
&t^R_{n\ell m} = \frac{k}{2}  \int_{-L}^L dy \,  e^{k|y|} \Big\{ 
h_n \big( f_{R\ell}^*  g_{R m}  + g_{R\ell}^*  f_{R m} \big) 
+ k_n \big( f_{R\ell}^* f_{R m}  - g_{R\ell}^* g_{R m} \big) \Big\} ,  \cr
\noalign{\kern 3pt}
&t^L_{n\ell m} =   -\frac{k}{2}  \int_{-L}^L dy \,  e^{k|y|} \Big\{ 
h_n \big( f_{L\ell}^*  g_{L m}  + g_{L\ell}^*  f_{L m} \big) 
+ k_n \big( f_{L \ell}^* f_{L m}  - g_{L \ell}^* g_{L m} \big) \Big\}  .
\label{RSgaugeCoupling1}
\end{align}
Note that $t^{R/L}_{n\ell m}$ depends on $\theta_H$, $z_L$  and $c$.
(It does not depend on $k$ or $m_\KK$.)

\begin{figure}[bth]
\centering
\includegraphics[height=40mm]{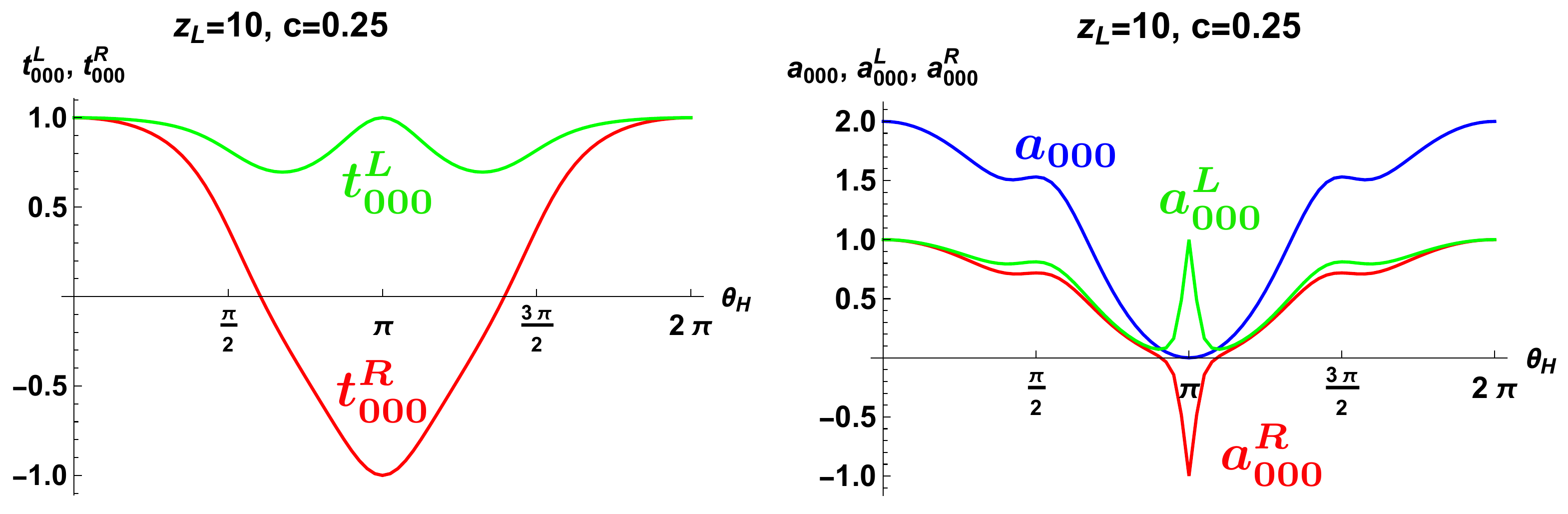}
\caption{Gauge couplings  $t^{R/L}_{000}$and anomaly coefficients
 $a^{R/L}_{000}$ and $a_{000}$  for $z_L=10$ and $c=0.25$  are shown
 as functions of $\theta_H$.     $a^{R/L}_{000}$ and $a_{000}$ are evaluated
 by taking account of $t^{R/L}_{0m \ell}$ ($0 \le m, \ell \le 14$).
}   
\label{fig:RSanomaly000}
\end{figure}

 Chiral anomaly in $\dd_\mu  j^\mu_{(n)}$ is written as 
 \begin{align}
&\dd_\mu  j^\mu_{(n)} + \cdots = -\Big( \frac{g_4}{2}\Big)^3 \sum_{\ell ,m =0}^\infty
\frac{a_{n\ell m}}{16\pi^2} \, Z_{\mu\nu}^{(\ell)} \tilde Z^{(m)\mu\nu}
\label{RSanomaly1}
\end{align}
where $Z_{\mu\nu}^{(\ell)} = \dd_\mu Z_\nu^{(\ell)}-  \dd_\nu Z_\mu^{(\ell)}$.
The anomaly coefficient $a_{n\ell m}$ is found to be
\begin{align}
a_{n_1 n_2 n_3} & = a_{n_1 n_2 n_3}^R +a_{n_1 n_2 n_3}^L ~,\cr
\noalign{\kern 3pt}
a_{n_1 n_2 n_3}^{R/L} &=
\sum_{m, \ell, p =0}^\infty  t_{n_1  m \ell}^{R/L} \,  t_{n_2 \ell p}^{R/L} \,  t_{n_3 p m}^{R/L} ~.
\label{RSanomaly2}
\end{align}
As the couplings  $t^{R}_{n\ell m}$ and  $t^{L}_{n\ell m}$ depend on $\theta_H$, 
$a^{R}_{n\ell m}$ and  $a^{L}_{n\ell m}$ also do depend on $\theta_H$.   In the RS space 
the dependence is smooth.  For instance, $t^{R}_{000}$,  $t^{L}_{000}$, $a^{R}_{000}$,  $a^{L}_{000}$
and  $a_{000}$ for $z_L=10$ and $c=0.25$  are  depicted in Fig.\  \ref{fig:RSanomaly000}.
It is seen that  $a_{000}$ changes from 2 to 0 as $\theta_H$ varies from 0 to $\pi$.

\section{Anomaly flow}

In the flat space the anomaly coefficients $b_{nm\ell}$ in (\ref{flatanomaly2}) are constant, whereas
the anomaly coefficients $a_{nm\ell}$ in (\ref{RSanomaly2}) in the RS space depend on $\theta_H$.
There is no contradiction between these two facts.    Look at the mass spectrum in Fig.\ \ref{fig:massspectrum}.
In the RS space the lowest mode of the gauge field is always $Z_\mu^{(0)}$ irrespective of the value of 
$\theta_H$.  In the flat space the lowest mode is $B_\mu^{(0)}$ for $0 < \theta_H < \onehalf \pi$, 
$B_\mu^{(-1)}$ for $ \onehalf \pi < \theta_H < \frac{3}{2}\pi$, and $B_\mu^{(-2)}$ for 
$ \frac{3}{2}\pi < \theta_H < 2 \pi$. The anomaly coefficient $b_{nnn}$ is $+2$ for  $n=0$ and $-2$,
but is 0 for $n=-1$.  As the AdS curvature approaches 0, that is, as $k \go 0$, the RS space
becomes the flat $M^4 \times (S^1/Z_2)$ space.  In other words, $a_{000}$ must flow 
from 2 to 0 to 2 as $\theta_H$ changes from 0 to $\pi$ to $2\pi$.
{\bf The anomaly flows as the AB phase $\theta_H$ varies.}
In the flat space the behavior of the anomaly becomes singular at the points of the level crossing,
namely at $\theta_H =  \onehalf \pi$ and $ \frac{3}{2}\pi$.

\begin{figure}[bth]
\centering
\includegraphics[height=40mm]{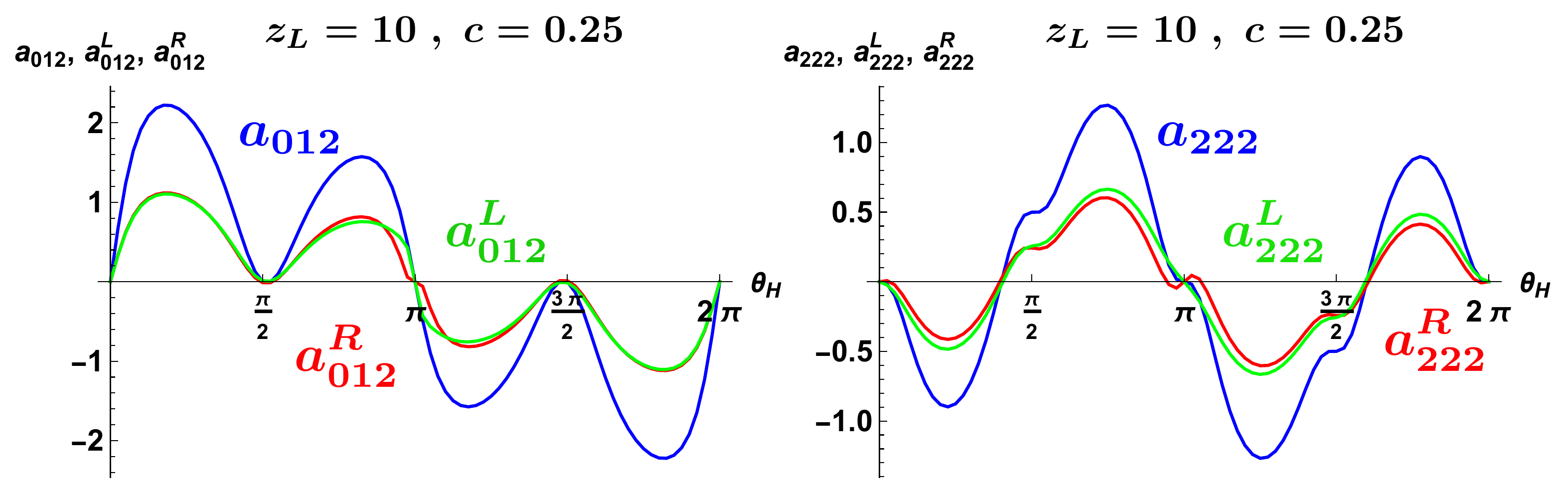}
\caption{Anomaly coefficients $a^{R/L}_{012}$, $a_{012}$,  $a^{R/L}_{222}$ and $a_{222}$
$a^{R/L}_{012}$, $a_{012}$,  $a^{R/L}_{222}$ and $a_{222}$  
for $z_L=10$ and $c=0.25$  are shown as functions of $\theta_H$.     
The coefficients are evaluated by taking account of $t^{R/L}_{jm \ell}$ ($j=0,1,2$, $0 \le m, \ell \le 14$).
}   
\label{fig:RSanomaly012}
\end{figure}

The phenomenon of the anomaly flow is seen in all anomaly coefficients $a_{n m \ell}$.
In Fig.\  \ref{fig:RSanomaly012}  the anomaly coefficients $a^{R/L}_{012}$, $a_{012}$,  
$a^{R/L}_{222}$ and $a_{222}$ are plotted  for $z_L=10$ and $c=0.25$.
The anomaly flow is smooth. 

\begin{figure}[bth]
\centering
\includegraphics[height=45mm]{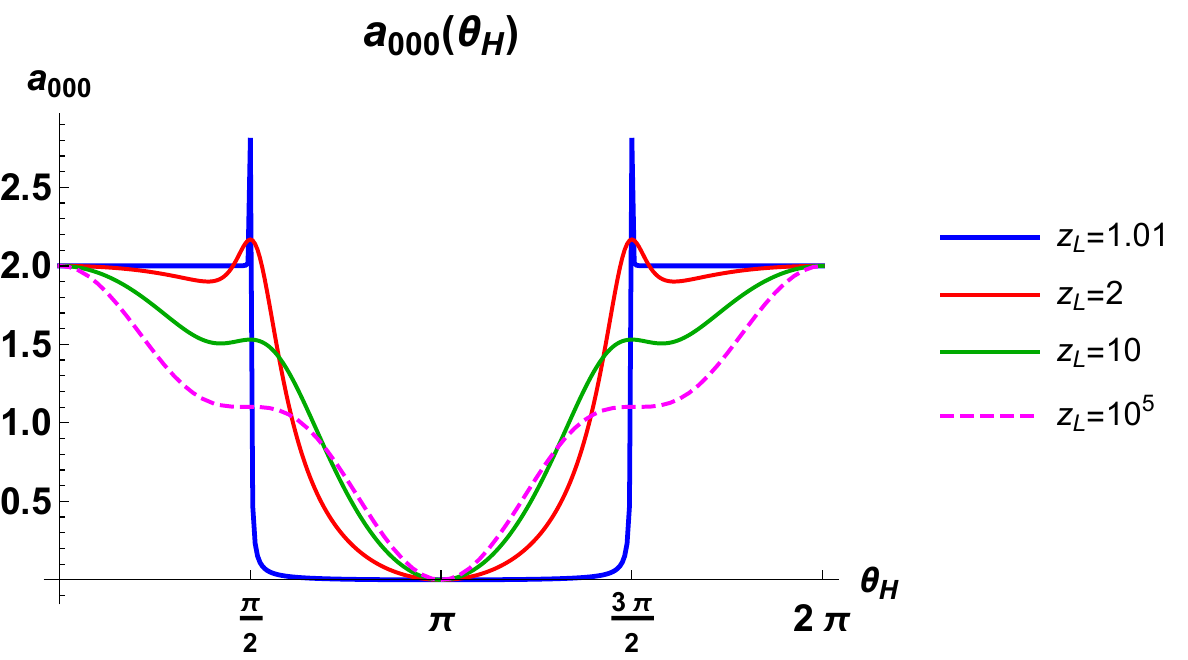}
\caption{The $z_L$-dependence of the anomaly coefficient $a_{000} (\theta_H; z_L)$
is displayed for $c=0.25$.
}   
\label{fig:anomaly000zLdep}
\end{figure}

One might wonder how the anomaly flow in the RS space reduces to the anomaly flow in the flat space
which seems singular at $\theta_H =  \onehalf \pi$ and $ \frac{3}{2}\pi$.
The flat space limit is obtained by taking  the $k \go 0$ limit in the RS space.   
As  $k \go 0$ with $L= \pi R$ kept fixed, $z_L \go 1$. 
In Fig.\ \ref{fig:anomaly000zLdep}  the behavior of $a_{000}(\theta_H; z_L)$ is shown.

One sees that $a_{000} (\theta_H; z_L)$ varies smoothly as $z_L$ changes.
The $z_L \go 1$ limit is singular at $\theta_H = \onehalf \pi $ and $\frac{3}{2} \pi$, however.
\begin{align}
&\lim_{z_L \go 1}  a_{000} (\theta_H; z_L) =
\begin{cases}
2 &{\rm for~} 0 \le \theta_H < \onehalf \pi \cr
2 \sqrt{2}  &{\rm for~}  \theta_H = \onehalf \pi \cr
0 &{\rm for~} \onehalf \pi  \le \theta_H < \frac{3}{2}  \pi \cr
2 \sqrt{2}  &{\rm for~} \theta_H = \frac{3}{2} \pi \cr
2 &{\rm for~}  \frac{3}{2} \pi  \le \theta_H \le 2 \pi 
\end{cases} ~.
\label{flatlimit1}
\end{align}
This is precisely the behavior in the flat space as
\begin{align}
&\lim_{z_L \go 1}  Z_\mu^{(0)} =
\begin{cases}
B_\mu^{(0)} &{\rm for~} 0 \le \theta_H < \onehalf \pi \cr
\frac{1}{\sqrt{2}} (B_\mu^{(0)}  + B_\mu^{(-1)} )  &{\rm for~}  \theta_H = \onehalf \pi \cr
B_\mu^{(-1)}&{\rm for~} \onehalf \pi  < \theta_H < \frac{3}{2}  \pi \cr
\frac{1}{\sqrt{2}} (B_\mu^{(-1)}  + B_\mu^{(-2)} )  &{\rm for~} \theta_H = \frac{3}{2} \pi \cr
B_\mu^{(-2)} &{\rm for~}  \frac{3}{2} \pi < \theta_H \le 2 \pi 
\end{cases} ~.
\label{flatlimit2}
\end{align}
As a function of $\theta_H$, the anomaly coefficient $a_{000}  (\theta_H)$ becomes singular
in the flat space limit at the points where level crossing occurs.

\section{Holography in anomaly flow}

In Figs.\ \ref{fig:RSanomaly000}, \ref{fig:RSanomaly012} and \ref{fig:anomaly000zLdep}, 
the anomaly coefficients coming from a fermion doublet of type 1A with the bulk mass 
parameter $c=0.25$ have been shown.  A surprise comes when one investigates the $c$-dependence of 
the anomaly coefficients $a_{nm\ell} (\theta_H; z_L, c)$.
In the realistic GHU models of electroweak interactions, namely in the $SO(5) \times U(1)_X \times SU(3)_C$
GHU in the RS space, \cite{GUTinspired2019} the top quark multiplet has $c \sim 0.3$ 
whereas the multiplets of other quarks and leptons  have $c= 0.6 \sim 1$.  

\begin{figure}[bth]
\centering
\includegraphics[height=40mm]{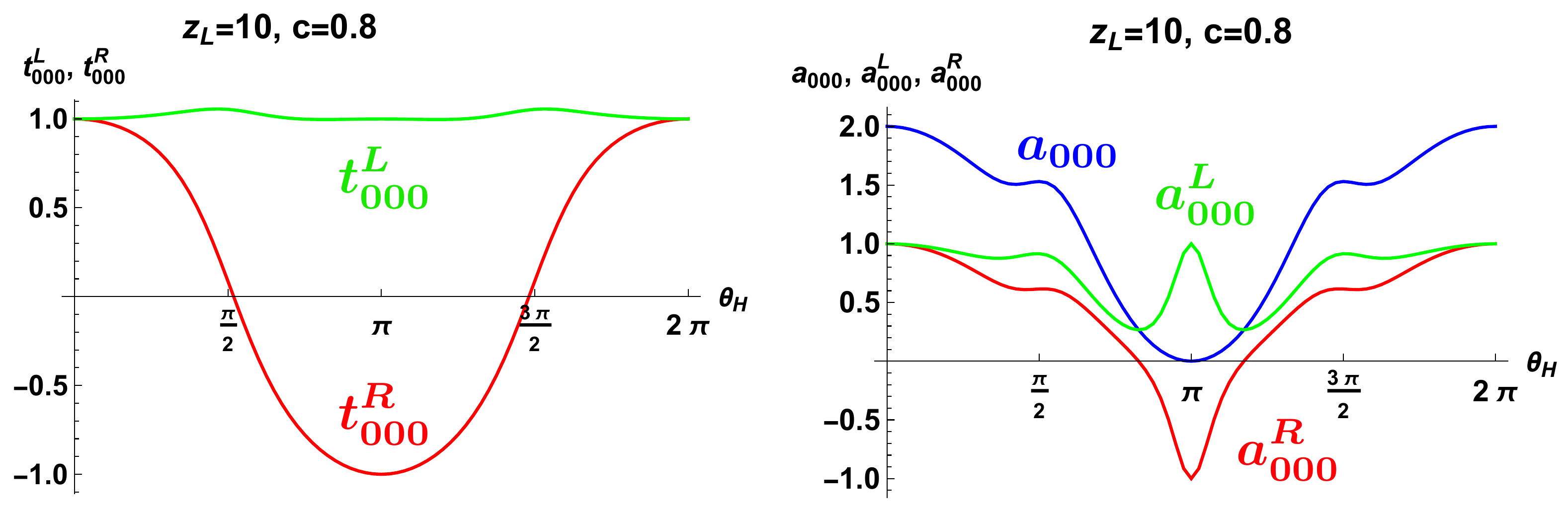}
\caption{Gauge couplings  $t^{R/L}_{000}$and anomaly coefficients
 $a^{R/L}_{000}$ and $a_{000}$  for $z_L=10$ and $c=0.8$  are shown
 as functions of $\theta_H$.  The anomaly coefficient $a_{000}$ for  $c=0.8$ has the same behavior
 as  $a_{000}$ for  $c=0.25$ in Fig.\ \ref{fig:RSanomaly000}.
}   
\label{fig:RSanomaly000c0.8}
\end{figure}

In Fig.\ \ref{fig:RSanomaly000c0.8} gauge couplings  $t^{R/L}_{000}$and anomaly coefficients
$a^{R/L}_{000}$ and $a_{000}$  for $z_L=10$ and $c=0.8$  are shown.   
The behavior in Fig.\ \ref{fig:RSanomaly000c0.8} should be compared with that in  in 
Fig.\ \ref{fig:RSanomaly000} for $c=0.25$.
Although the gauge couplings for $c=0.8$  are significantly different from those for $c=0.25$,
the total anomaly coefficient $a_{000} (\theta_H)$ turns out universal, being independent of the 
value of $c$.   There must be a reason for this fact.

Let us go back to the expression for $a_{n_1 n_2 n_3}$ in (\ref{RSanomaly2}) with (\ref{RSgaugeCoupling1}).
\begin{align}
&a_{n_1 n_2 n_3} = \sum_{m, \ell , p} \Big\{ t_{n_1  m \ell}^{R} \,  t_{n_2 \ell p}^{R} \,  t_{n_3 p m}^{R} 
+ t_{n_1  m \ell}^{L} \,  t_{n_2 \ell p}^{L} \,  t_{n_3 p m}^{L}  \Big\} ~.
\label{RSanomaly3}
\end{align}
There are two ways to evaluate $a_{n_1 n_2 n_3}$.

\noindent  ~~ 
{\bf Method 1} ~  (i) First evaluate the couplings $t_{nm\ell}^{R/L}\,$.  (ii) Then do the summation $\sum_{m \ell p}$.

\noindent ~~ 
{\bf Method 2} ~  (i) First do the summation $\sum_{m \ell p}$.   (ii) Then do the integration $\int dy_1 dy_2 dy_3$.

\noindent
So far we have adopted Method 1 to evaluate $a_{n_1 n_2 n_3}$. 

In Method 2 the first step of  the summation  $\sum_{m \ell p}$ leads to
\begin{align}
&a_{n_{1} n_{2} n_{3}} = \Big( \frac{k }{2} \Big)^{3} \int \int \int_{-\beta}^{2L-\beta} dy_1 dy_2 dy_3 \,
e^{\sigma (y_1) + \sigma (y_2) + \sigma (y_3)} \cr
\noalign{\kern 5pt}
&\quad
\times \Big[ ~  k_1 k_2 k_3 \big\{ A_R(1,2) A_R(2,3) A_R (3,1) - B_R(1,2) B_R(2,3) B_R (3,1)  \cr
\noalign{\kern 2pt}
&\hskip 2.2cm
+ B_L(1,2) B_L(2,3) B_L (3,1) - A_L(1,2) A_L(2,3) A_L (3,1) \big\} \cr
\noalign{\kern 2pt}
&\qquad
+ k_1 h_2 h_3 \{ \cdots \} + h_1 k_2 h_3 \{ \cdots \} + h_1 h_2 k_3 \{ \cdots \} \Big] ~, \cr
\noalign{\kern 3pt}
&\quad
k_j = k_{n_j} (y_j) ~, ~~ h_j = h_{n_j} (y_j) ~, \cr
\noalign{\kern 5pt}
&\quad
\begin{pmatrix} A_{R/L} (j, k) \cr B_{R/L} (j, k) \end{pmatrix}
 = \begin{pmatrix} A_{R/L} \cr B_{R/L}  \end{pmatrix}  (y_j, y_k) 
 = \sum_{n=0}^\infty  \begin{pmatrix}  f_{R/L n} (y_j) f_{R/L n}^* (y_k)  \cr
g_{R/L n} (y_j) g_{R/L n}^* (y_k)  \end{pmatrix} .
\label{RSanomaly4}
\end{align}
Here we have made use of $\sum_{n=0}^\infty f_{R/L n} (y) g_{R/L n}^* (y') =0$.
$\beta$ appearing in the integration range in $y_j$ is arbitrary.
Finding the explicit form of $A_{R/L}$ and $B_{R/L}$ for general $c$ is a difficult task, however.

It is possible to determine $A_{R/L}$ and $B_{R/L}$ for $c=0$.   One finds, with  $\delta_L (x) = \sum_n \delta(x - nL)$, that
\begin{align}
&\underline{\hbox{type 1A}, ~ c=0} \cr
\noalign{\kern 2pt}
&A_R(y,y') = B_L (y,y')
= \frac{e^{-\sigma (y)}}{k} \big\{ \delta_{2L} (y -y') + \delta_{2L} (y +y')  \big\} ~, \cr
\noalign{\kern 2pt}
&B_R(y,y') = A_L (y,y') 
= \frac{e^{-\sigma (y)}}{k} \big\{ \delta_{2L} (y -y')  - \delta_{2L} (y +y') \big\} ~,
\label{completeness1}
\end{align}
Formulas for type 1B are obtained by interchanging $R$ and $L$.   Similarly
\begin{align}
&\underline{\hbox{type 2A}, ~ c=0} \cr
\noalign{\kern 2pt}
&A_R(y,y') = B_L (y,y')
= \frac{e^{-\sigma (y)}}{k} \big\{ \hat \delta_{2L} (y -y') + \hat \delta_{2L} (y +y')  \big\} ~, \cr
\noalign{\kern 2pt}
&B_R(y,y') = A_L (y,y') 
= \frac{e^{-\sigma (y)}}{k} \big\{ \hat \delta_{2L} (y -y')  - \hat \delta_{2L} (y +y') \big\} ~, \cr
\noalign{\kern 4pt}
&\hat \delta_{2L} (y) =   \delta_{4L} (y) - \delta_{4L} (y - 2L) ~.
\label{completeness2}
\end{align}
Formulas for type 2B are obtained by interchanging $R$ and $L$.  When one inserts (\ref{completeness1}) and
(\ref{completeness2}) into (\ref{RSanomaly4}), there appear the products of three delta functions in the integrand.
With $0 < \beta < L$ the products of delta functions reduce to 
\begin{align}
&\begin{matrix} \delta_{2L} (y_1 - y_2)  \delta_{2L} (y_2 - y_3)  \delta_{2L} (y_3 + y_1) \cr
\noalign{\kern 3pt}
\delta_{2L} (y_1 + y_2)  \delta_{2L} (y_2 +y_3)  \delta_{2L} (y_3 + y_1) \end{matrix} ~ \bigg\} \cr
\noalign{\kern 3pt}
&\quad
\Rightarrow \frac{1}{2} \Big\{ \delta (y_1) \delta (y_2) \delta (y_3) + \delta (y_1 -L) \delta (y_2-L) \delta (y_3 -L) \Big\}  ~, \cr
\noalign{\kern 3pt}
&\begin{matrix} \hat \delta_{2L} (y_1 - y_2)  \hat \delta_{2L} (y_2 - y_3)  \hat \delta_{2L} (y_3 + y_1) \cr
\noalign{\kern 3pt}
\hat \delta_{2L} (y_1 + y_2)  \hat \delta_{2L} (y_2 +y_3)  \hat \delta_{2L} (y_3 + y_1) \end{matrix} ~ \bigg\} \cr
\noalign{\kern 3pt}
&\quad 
\Rightarrow \frac{1}{2} \Big\{ \delta (y_1) \delta (y_2) \delta (y_3) - \delta (y_1 -L) \delta (y_2-L) \delta (y_3 -L) \Big\}  ~.
\label{deltaFn1}
\end{align}
Furthermore, as $h_n (0) = h_n (L) = 0$, only the terms proportional to $k_1 k_2 k_3$ in (\ref{RSanomaly4}) survive.

We have arrived at the following formula for the anomaly coefficients;
\begin{align}
a_{n \ell m} (\theta_H, z_L) &=  Q_0 k_n (0) k_\ell (0) k_m (0) + Q_1 k_n (L) k_\ell (L) k_m (L)  ~, 
\label{AnomalyFormula1}
\end{align}
where
\begin{align}
(Q_0, Q_1) &= \begin{cases} (+1, +1) &\hbox{for type 1A} \cr   (-1, -1) &\hbox{for type 1B} \cr
 (+1, -1) &\hbox{for type 2A} \cr  (-1, +1) &\hbox{for type 2B} \end{cases} ~.
\label{AnomalyFormula2}
\end{align}
The anomaly coefficients are determined by the values of the wave functions of the gauge fields, $k_n(y)$,  
at the UV and IR branes and the parity, $Q_j$,  of the right-handed mode of the fermion field.

\begin{figure}[bth]
\centering
\includegraphics[height=80mm]{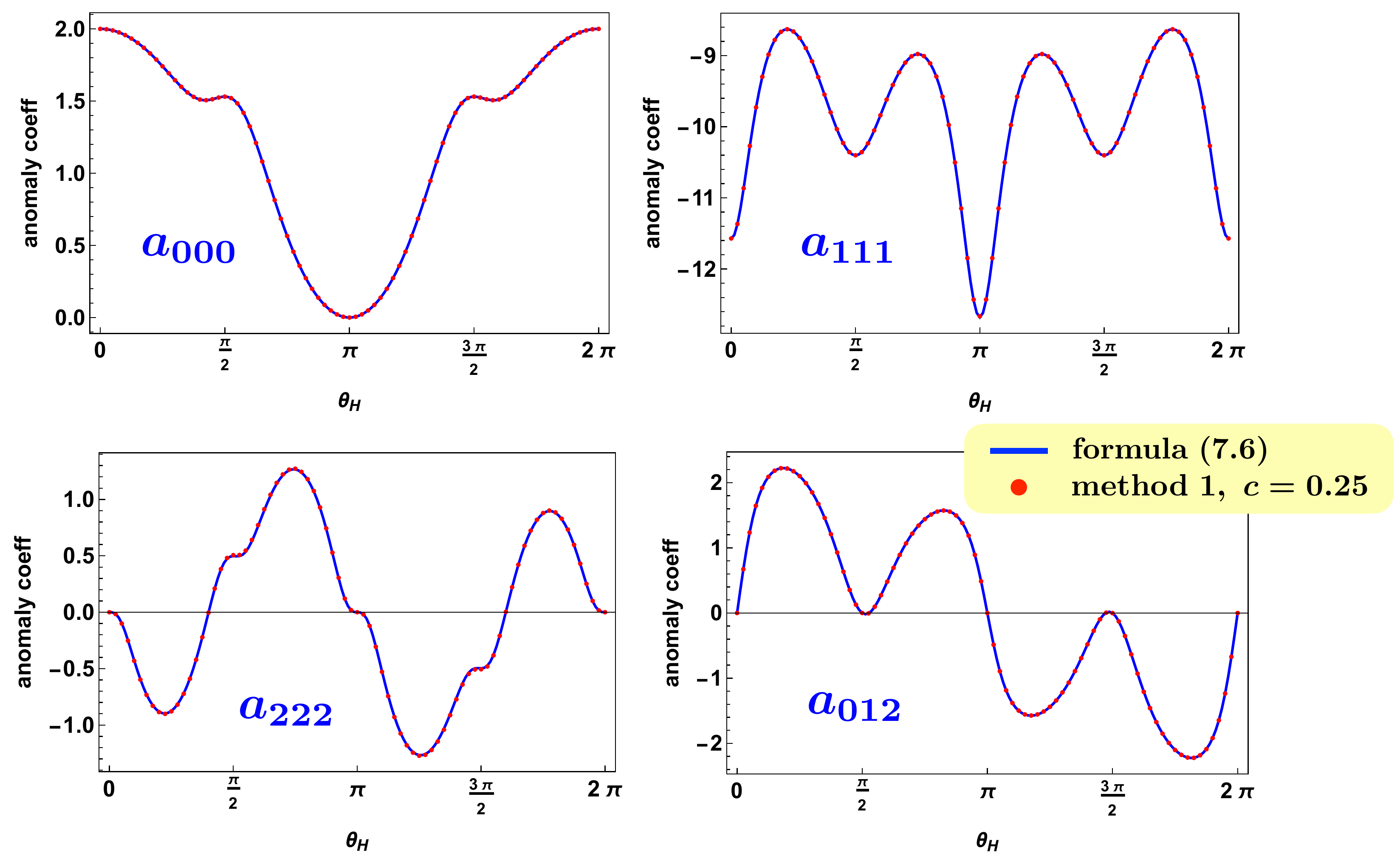}
\caption{Anomaly coefficients $a_{000} (\theta_H)$,  $a_{111} (\theta_H)$ ,  $a_{222} (\theta_H)$ ,  $a_{012} (\theta_H)$ 
are shown for type 1A fermions with $z_L=10$.
Blue curves represent the formula  (\ref{AnomalyFormula1}), whereas red dots represent the values determined
from the gauge couplings $t_{nm\ell}^{R/L}$ for $c=0.25$ by Method 1.
}   
\label{fig:anomalyformula}
\end{figure}

As observed at the beginning of this section, the anomaly coefficients $a_{n \ell m} (\theta_H)$ do not depend on
the bulk mass parameter $c$ of the fermion field.  The anomaly formula (\ref{AnomalyFormula1})  derived for 
$c=0$ should apply for other values of $c$.  Indeed one can confirm it explicitly.
In Fig.\  \ref{fig:anomalyformula} $a_{n \ell m}$'s given by the formula  (\ref{AnomalyFormula1}) and those determined by Method 1
for $c=0.25$ are displayed in the case $z_L=10$.
Blue curves represent the formula  (\ref{AnomalyFormula1}), whereas red dots represent the values determined
from the gauge couplings $t_{nm\ell}^{R/L}$ for $c=0.25$.  It is seen that the red dots are on the blue curves.

We stress that the anomaly coefficients  depend solely on $k_n(0)$, $k_n(L)$, $Q_0$ and $Q_1$.
They depend neither on the behavior of the wave functions of gauge fields and fermion field in the bulk region $0 < y < L$,
nor on the bulk mass parameter $c$ of the fermion field.  The formula  (\ref{AnomalyFormula1}) represents
{\bf holography in anomaly flow}.   Although each anomaly generated by specific fermion modes running along 
the internal triangle loop does depend on the detailed behavior of the wave functions of both gauge 
and fermion fields in the bulk, the total anomaly coefficients, after summing up all possible loop contributions, 
are determined by the information of the gauge fields at the UV and IR branes and of the parity conditions for
the fermion field  there.

The formula for the anomaly coefficients $b_{nm\ell}$ in the flat $M^4 \times (S^1/Z_2)$ spacetime simplifies.
As in the RS space, one finds that
\begin{align}
b_{n \ell m}  &=  Q_0 k^{\rm flat}_n (0) k^{\rm flat}_\ell (0) k^{\rm flat}_m (0) 
+ Q_1 k^{\rm flat}_n (L) k^{\rm flat}_\ell (L) k^{\rm flat}_m (L)  ~.
\label{flatAnomaly3}
\end{align}
In the flat space $k_n^{\rm flat} (y) = \cos (n\pi y/L)$ so that
\begin{align}
b_{n \ell m}  &=  Q_0 + (-1)^{n + \ell + m}  \,  Q_1 ~,
\label{flatAnomaly4}
\end{align}
which agrees with (\ref{flatanomaly2}) with $Q_0=Q_1 =1$ for a doublet fermion of type 1A.

\section{Anomaly cancellation}

In gauge theory in four dimensions gauge anomalies have to be cancelled for the consistency
of the theory.  In GHU the holography in anomaly flow becomes crucial to guarantee the cancellation
of gauge anomalies.  In the SM all gauge anomalies are cancelled among quarks and leptons in each
generation.\cite{Bouchiat1972, GrossJackiw1972}

In the $SU(2)$ GHU under consideration one may introduce several fermion doublets, each of which 
has its own bulk mass parameter $c$.   Let the numbers of doublet fermions
of type 1A, 1B, 2A and 2B be $n_{1A}$, $n_{1B}$, $n_{2A}$ and $n_{2B}$, respectively.  
It follows from (\ref{AnomalyFormula1})  that the anomalies are cancelled  if 
\begin{align}
&n_{1A} = n_{1B}  ~,~~ n_{2A} = n_{2B}  ~,
\label{cancellation1}
\end{align}
as the anomaly coefficients do not dependon $c$.
The condition is generalized in the presence of brane fermions, namely fermions living only on the UV or IR brane.
 Suppose that there are $\hat n_R$ right-handed and  $\hat n_L$ left-handed  doublet brane fermions 
on the UV brane at $y=0$.
As the $Z_\mu^{(n)}$ coupling of each brane fermion is given by $(g_4/2) \, k_n(0)$, the anomaly
cancellation conditions become
\begin{align}
&n_{1A} - n_{1B}  + n_{2A} - n_{2B}  + \hat n_R - \hat n_L = 0 ~, \cr
&n_{1A} - n_{1B}  - n_{2A} + n_{2B} = 0 ~,
\label{cancellation2}
\end{align}
It is important that  the conditions (\ref{cancellation1}) and (\ref{cancellation2}) do not depend on $\theta_H$ and $z_L$.
The conditions guarantee that not only the zero mode anomaly $a_{000}$ but also all other anomalies 
$a_{n\ell m}$ are cancelled at once.
In GHU in the RS space the gauge couplings vary as $\theta_H$, which further depends on $c$, or on the fermion species,
but the anomaly cancellation conditions do not depend on $\theta_H$.

\section{Summary}

We have shown that the anomaly flows with the AB phase $\theta_H$.  In the RS space everything changes
smoothly with $\theta_H$.   In the $SU(2)$ GHU model in the RS space a chiral fermion at $\theta_H=0$ is
transformed to a vector-like fermion at $\theta_H = \pi$.   
The magnitude of the anomaly coming from one fermion doublet varies with $\theta_H$.
The total anomaly coefficients $a_{nm\ell} (\theta_H)$ are given by thec formula  (\ref{AnomalyFormula1}),
which represents holography in anomaly flow.

The anomalies can be cancelled among several fermion doublets.  The cancellation conditions are given by
(\ref{cancellation1}) or (\ref{cancellation2}).   They are independent of $\theta_H$ and $z_L$, which is
important to achieve the anomaly cancellation in the realistic GHU models in the RS space.
The examination of  anomaly cancellation in the $SO(5) \times U(1)_X \times SU(3)_C$  GHU in the RS
space is necessary.  

In this connection one may worry about the fact that the gauge couplings vary with $\theta_H$, and
the couplings of quarks and leptons are not purely chiral at $\theta_H \not= 0$.
In the SM only left-handed quark-lepton doublets couple to $W$.    In GHU models in the RS space 
right-handed components also have small  couplings to $W$ at $\theta_H \not= 0$.
The detailed study of the GUT inspired  $SO(5) \times U(1)_X \times SU(3)_C$  GHU has been done
recently.\cite{CouplingSumRule}. In the realistic model $\theta_H \sim0.1$, $m_\KK \sim 13\,$TeV 
and $z_L \sim 4 \times 10^{11}$.   The $W$ couplings of right-handed quarks in units of $g_w$ are 
$O(10^{-12})$, $O(10^{-9})$, and $O(10^{-5})$ for $(u,d)$, $(c,s)$ and $(t,b)$, respectively.
The $W$ couplings of  right-handed leptons are much smaller.

Anomaly flow by an Aharonov-Bohm phase is a new phenomenon, which is different from
the phenomenon of anomaly inflow.\cite{CallanHarvey1985, Fukaya2017, WittenYonekura2021}
Further investigation is desired.

\section*{Acknowledgement}

This work was supported in part by Japan Society for the Promotion of Science, Grants-in-Aid for Scientific 
Research, Grant No. JP19K03873.

\vfill

\appendix

\section{Basis functions} 

Wave functions of gauge fields and fermions in the RS space are expressed in terms of the Bessel functions.
For gauge fields we introduce 
\begin{align}
 C(z; \lambda) &= \frac{\pi}{2} \lambda z z_L F_{1,0}(\lambda z, \lambda z_L) ~,  \cr
 S(z; \lambda) &= -\frac{\pi}{2} \lambda  z F_{1,1}(\lambda z, \lambda z_L) ~, \cr
 C^\prime (z; \lambda) &= \frac{\pi}{2} \lambda^2 z z_L F_{0,0}(\lambda z, \lambda z_L) ~,  \cr
S^\prime (z; \lambda) &= -\frac{\pi}{2} \lambda^2 z  F_{0,1}(\lambda z, \lambda z_L)~, \cr
\noalign{\kern 5pt}
 F_{\alpha, \beta}(u, v) &\equiv 
J_\alpha(u) Y_\beta(v) - Y_\alpha(u) J_\beta(v) ~,
\label{functionA1}
\end{align}
where $J_\alpha (u)$ and $Y_\alpha (u)$ are Bessel functions of  the first and second kind.
These functions satisfy
\begin{align}
&- z \frac{d}{dz} \frac{1}{z} \frac{d}{dz} \begin{pmatrix} C \cr S \end{pmatrix} 
= \lambda^{2} \begin{pmatrix} C \cr S \end{pmatrix} ~, ~~
- \frac{d}{dz} z \frac{d}{dz} \frac{1}{z} \begin{pmatrix} C' \cr S' \end{pmatrix} 
= \lambda^{2} \begin{pmatrix} C' \cr S' \end{pmatrix} ~,   \cr
\noalign{\kern 5pt}
&C(z_{L} ; \lambda)  = z_{L} ~, ~~ C' (z_{L} ; \lambda)  = 0 ~, 
~~ S(z_{L} ; \lambda)  = 0 ~, ~~ S' (z_{L} ; \lambda)  = \lambda ~,  \cr
\noalign{\kern 5pt}
&CS' - S C' = \lambda z ~.
\label{relationA1}
\end{align}
For fermion fields with a bulk mass parameter $c$, we define 
\begin{align}
\begin{pmatrix} C_L \cr S_L \end{pmatrix} (z; \lambda,c)
&= \pm \frac{\pi}{2} \lambda \sqrt{z z_L} F_{c+\frac12, c\mp\frac12}(\lambda z, \lambda z_L) ~, \cr
\begin{pmatrix} C_R \cr S_R \end{pmatrix} (z; \lambda,c)
&= \mp \frac{\pi}{2} \lambda \sqrt{z z_L} F_{c- \frac12, c\pm\frac12}(\lambda z, \lambda z_L) ~.
\label{functionA3}
\end{align}
These functions satisfy 
\begin{align}
&D_{+} (c) \begin{pmatrix} C_{L} \cr S_{L} \end{pmatrix} = \lambda  \begin{pmatrix} S_{R} \cr C_{R} \end{pmatrix}, \cr
\noalign{\kern 5pt}
&D_{-} (c) \begin{pmatrix} C_{R} \cr S_{R} \end{pmatrix} = \lambda  \begin{pmatrix} S_{L} \cr C_{L} \end{pmatrix}, ~~
D_{\pm} (c) = \pm \frac{d}{dz} + \frac{c}{z} ~, \cr
\noalign{\kern 5pt}
&C_{R} = C_{L} = 1 ~, ~~ S_{R} = S_{L} = 0 \quad {\rm at~} z=z_{L} ~, \cr
\noalign{\kern 5pt}
&C_L C_R - S_L S_R=1 ~. 
\label{relationA2}
\end{align}

\section{Wave functions in RS} 

The wave functions $\tilde {\bf h}_n(z) \equiv ( \tilde h_n(z) ,  \tilde k_n(z) )^t$ in (\ref{RSgaugeKK1})
for gauge fields  are  given by
\begin{align}
&\tilde{\bf h}_0 (z) = \bar {\bf h}_0^a (z) ~, \cr
\noalign{\kern 3pt}
&[ \tilde{\bf h}_{2\ell - 1} (z) ,  \tilde{\bf h}_{2\ell} (z)] = (-1)^\ell \begin{cases} 
[\bar {\bf h}_{2\ell - 1}^a (z), \bar {\bf h}_{2\ell }^b (z) ] &{\rm for} - \frac{1}{2} \pi  <  \theta_H < \frac{1}{2} \pi \cr
[\bar {\bf h}_{2\ell - 1}^b (z) , - \bar {\bf h}_{2\ell}^a (z) ]&{\rm for~} 0 < \theta_H < \pi \cr
[- \bar {\bf h}_{2\ell - 1}^a (z), - \bar {\bf h}_{2\ell}^b (z)] &{\rm for~}  \frac{1}{2} \pi < \theta_H < \frac{3}{2} \pi \cr
[-\bar {\bf h}_{2\ell - 1}^b (z), \bar {\bf h}_{2\ell}^a (z)] &{\rm for~}  \pi < \theta_H < 2 \pi \cr
[\bar {\bf h}_{2\ell - 1}^a (z) , \bar {\bf h}_{2\ell }^b (z)]&{\rm for~} \frac{3}{2} \pi <  \theta_H <  \frac{5}{2} \pi 
\end{cases} 
\cr
\noalign{\kern 3pt}
&\hskip 3.5cm
(\ell = 1, 2, 3, \cdots) ~, \cr
\noalign{\kern 5pt}
&\bar {\bf h}_n^a (z) 
= \frac{1}{\sqrt{r^a_n}} \begin{pmatrix} - \sin \theta_H  \hat S (z; \lambda_n) \cr \cos \theta_H C (z; \lambda_n)   \end{pmatrix} , ~~
\bar {\bf h}_n^b (z)  
= \frac{1}{\sqrt{r^b_n}} \begin{pmatrix}\cos \theta_H   S (z; \lambda_n) \cr \sin \theta_H \check C (z; \lambda_n)   \end{pmatrix} , \cr
\noalign{\kern 3pt}
&\hskip 1.cm
\hat S (z; \lambda) =  \frac{C(1;\lambda)}{S(1;\lambda)} \, S (z; \lambda) ~,~~
\check C (z; \lambda) =  \frac{S'(1;\lambda)} {C'(1;\lambda)} \, C (z; \lambda)  ~, \cr
\noalign{\kern 3pt}
&\hskip 1.cm
r_{n} = \frac{1}{kL} \int_{1}^{z_{L}} \frac{dz}{z} \big\{ | \hat h_{n} (z) |^{2} + | \hat k_{n} (z) |^{2} \big\}
\quad {\rm for}~ \begin{pmatrix} \hat h_{n} (z) \cr \hat k_{n} (z) \end{pmatrix} .
\label{RSgaugeKKB1}
\end{align}
In (\ref{RSgaugeKKB1}) two expressions in an overlapping region in $\theta_H$ are the same.

The wave functions $\tilde{\bf f}_{Rn} (z) = (\tilde f_{Rn} (z), \tilde g_{Rn} (z) )^t$ and
$\tilde{\bf f}_{L n} (z) = (\tilde f_{L n} (z), \tilde g_{L n} (z) )^t$ in (\ref{RSfermionKK2})
for fermion fields  of type 1A with $c\ge 0$ are  given by
\begin{align}
[\tilde{\bf f}_{R, 2\ell} (z) , \tilde{\bf f}_{R, 2\ell +1} (z) ] &= 
\begin{cases} [\bar {\bf f}_{R, 2\ell}^a (z),  \bar {\bf f}_{R, 2\ell +1}^c (z)] &{\rm for~} - \pi < \theta_H < \pi  \cr
[\bar {\bf f}_{R, 2\ell}^b (z) ,  \bar {\bf f}_{R, 2\ell +1}^d (z)] &{\rm for~} 0 < \theta_H <  2\pi \cr
[- \bar {\bf f}_{R, 2\ell}^a (z), - \bar {\bf f}_{R, 2\ell +1}^c (z)] &{\rm for~}   \pi < \theta_H < 3\pi  \cr
[- \bar {\bf f}_{R, 2\ell}^b (z) , - \bar {\bf f}_{R, 2\ell +1}^d (z)]&{\rm for~} 2 \pi < \theta_H <  4\pi \cr
[\bar {\bf f}_{R, 2\ell}^a (z) , \bar {\bf f}_{R, 2\ell +1}^c (z)]&{\rm for~}  3 \pi < \theta_H < 5 \pi  \cr \end{cases}  \cr
&\hskip 1.cm  
(\ell=0,1,2, \cdots) ~, \cr
\noalign{\kern 5pt}
\tilde{\bf f}_{L0} (z)  &=  \bar {\bf f}_{L0}^a (z) , \cr
\noalign{\kern 5pt}
[\tilde{\bf f}_{L, 2\ell-1} (z), \tilde{\bf f}_{L, 2\ell} (z) ]  &= 
\begin{cases} [\bar {\bf f}_{L, 2\ell-1}^a (z) ,  \bar {\bf f}_{L, 2\ell}^c (z)] &{\rm for~} - \pi < \theta_H < \pi  \cr
[\bar {\bf f}_{L, 2\ell-1}^b (z) ,  \bar {\bf f}_{L, 2\ell}^d (z)]&{\rm for~} 0 < \theta_H <  2\pi \cr
[- \bar {\bf f}_{L, 2\ell-1}^a (z) , - \bar {\bf f}_{L, 2\ell}^c (z)]&{\rm for~}   \pi < \theta_H < 3\pi  \cr
[- \bar {\bf f}_{L, 2\ell-1}^b (z) , - \bar {\bf f}_{L, 2\ell}^d (z)] &{\rm for~} 2 \pi < \theta_H <  4\pi \cr
[\bar {\bf f}_{L, 2\ell-1}^a (z) ,  \bar {\bf f}_{L, 2\ell}^c (z)]&{\rm for~}  3 \pi < \theta_H < 5 \pi  \cr \end{cases}  \cr
&\hskip 1.cm  
(\ell=1,2, 3,  \cdots) ~.
\label{RSfermionKKB1}
\end{align}
Here
\begin{align}
&\bar {\bf f}_{Rn}^a (z) = \frac{1}{\sqrt{r^a_n}} 
\begin{pmatrix}  \cos \onehalf \theta_H C_{R}(z; \lambda_{n}, c) \cr -  \sin \onehalf \theta_H \hat S_{R} (z; \lambda_{n}, c)  \end{pmatrix} , ~~
\bar {\bf f}_{Rn}^b (z)  = \frac{1}{\sqrt{r^b_n}} 
\begin{pmatrix}  \sin \onehalf \theta_H C_{R}(z; \lambda_{n}, c) \cr  \cos \onehalf \theta_H \check S_{R} (z; \lambda_{n}, c)  \end{pmatrix} ,  \cr
\noalign{\kern 5pt}
&\bar {\bf f}_{Rn}^c (z)  = \frac{1}{\sqrt{r^c_n}} 
\begin{pmatrix}  \sin \onehalf \theta_H \hat C_{R}(z; \lambda_{n}, c) \cr  \cos \onehalf \theta_H S_{R} (z; \lambda_{n}, c)  \end{pmatrix} , ~~
\bar {\bf f}_{Rn}^d (z)  = \frac{1}{\sqrt{r^d_n}} 
\begin{pmatrix}  - \cos \onehalf \theta_H \check C_{R}(z; \lambda_{n}, c) \cr  \sin \onehalf \theta_H S_{R} (z; \lambda_{n}, c)  \end{pmatrix} ,\cr
\noalign{\kern 5pt}
&\bar {\bf f}_{Ln}^a (z)  = \frac{1}{\sqrt{r^a_n}} 
\begin{pmatrix}  \sin \onehalf \theta_H  \hat S_{L}(z; \lambda_{n}, c) \cr \cos \onehalf \theta_H C_{L} (z; \lambda_{n}, c) \end{pmatrix} , ~~
\bar {\bf f}_{Ln}^b (z)  = \frac{1}{\sqrt{r^b_n}} 
\begin{pmatrix} -  \cos \onehalf \theta_H \check S_{L}(z; \lambda_{n}, c) \cr \sin \onehalf \theta_H C_{L} (z; \lambda_{n}, c) \end{pmatrix} , \cr
\noalign{\kern 5pt}
&\bar {\bf f}_{Ln}^c (z)  = \frac{1}{\sqrt{r^c_n}} 
\begin{pmatrix} \cos \onehalf \theta_H S_{L}(z; \lambda_{n}, c) \cr  -  \sin \onehalf \theta_H \hat C_{L} (z; \lambda_{n}, c) \end{pmatrix} , ~~
\bar {\bf f}_{Ln}^d (z)  = \frac{1}{\sqrt{r^d_n}} 
\begin{pmatrix}  \sin \onehalf \theta_H S_{L}(z; \lambda_{n}, c) \cr  \cos \onehalf \theta_H \check C_{L} (z; \lambda_{n}, c) \end{pmatrix} , \cr
\noalign{\kern 5pt}
&\qquad  r_n = \int_1^{z_L} dz  \big\{ | \hat f_{n} (z) |^{2} + | \hat g_{n} (z) |^{2} \big\} 
\quad {\rm for}~ \begin{pmatrix} \hat f_{n} (z) \cr \hat g_{n} (z) \end{pmatrix} ,
\label{RSfermionKKB2}
\end{align}
and
\begin{align}
&\begin{pmatrix} \hat S_{L} \cr \hat C_R \end{pmatrix} (z; \lambda,c) 
= \frac{C_{L} (1; \lambda,c)}{S_{L} (1; \lambda,c)} \begin{pmatrix} S_{L} \cr C_R \end{pmatrix}  (z; \lambda,c) , ~~
\begin{pmatrix} \hat C_{L} \cr \hat S_R \end{pmatrix} (z; \lambda,c) 
=  \frac{C_{R} (1; \lambda,c)}{S_{R} (1; \lambda,c)}  \begin{pmatrix} C_{L} \cr  S_R \end{pmatrix} (z; \lambda,c) ,  \cr
\noalign{\kern 5pt}
&\begin{pmatrix}\check S_{L} \cr \check C_R \end{pmatrix} (z; \lambda,c) 
= \frac{S_{R} (1; \lambda,c)}{C_{R} (1; \lambda,c)}  \begin{pmatrix}  S_{L} \cr C_R \end{pmatrix} (z; \lambda,c) , ~~
\begin{pmatrix} \check C_{L} \cr \check S_R \end{pmatrix} (z; \lambda,c) 
= \frac{S_{L} (1; \lambda,c)}{C_{L} (1; \lambda,c)}  \begin{pmatrix} C_{L} \cr S_R \end{pmatrix}   (z; \lambda,c) .
\label{functionB4}
\end{align}
In (\ref{RSfermionKKB1}) two expressions in an overlapping region in $\theta_H$ are the same.

\def\jnl#1#2#3#4{{#1}{\bf #2},  #3 (#4)}

\def\PRD{{\em Phys.\ Rev.} D}
\def\PRL{\em Phys.\ Rev.\ Lett. }
\def\PR{{\em Phys.\ Rev.} }

\def\PTEP{{\em Prog.\ Theoret.\ Exp.\  Phys.\  }}
\def\NP{{\em Nucl.\ Phys.} }
\def\NPB{{\em Nucl.\ Phys.} B}

\def\PLB{{\it Phys.\ Lett.} B}
\def\AP{{\em Ann.\ Phys.\ (N.Y.)} }
\def\NCA{{\it Nuovo Cim.} A}
\def\MPLA{{\em Mod.\ Phys.\ Lett.} A}


\renewenvironment{thebibliography}[1]
         {\begin{list}{[$\,$\arabic{enumi}$\,$]}  
         {\usecounter{enumi}\setlength{\parsep}{0pt}
          \setlength{\itemsep}{0pt}  \renewcommand{\baselinestretch}{1.2}
          \settowidth
         {\labelwidth}{#1 ~ ~}\sloppy}}{\end{list}}



\vskip 1.cm

\leftline{\large \bf References}

\begin{thebibliography}{99}


\bibitem{Adler1969}
S.L.\ Adler,
{\it ``Axial vector vertex in spinor electrodynamics''}, 
\jnl{\PR}{177}{2426}{1969}.

\bibitem{BellJackiw1969}
J.S.\ Bell and R.\ Jackiw,
{\it ``A PCAC puzzle: $\pi^0 \go \gamma \gamma$ in the $\sigma$ model''}, 
\jnl{\NCA}{60}{47}{1969}.

\bibitem{Fujikawa1979}
K.\ Fujikawa,
{\it ``Path-integral measure for gauge-invariant fermion theories''}, 
\jnl{\PRL}{42}{1195}{1979}; 
%
{\it ``Path integral  for gauge theories with fermions''}, 
\jnl{\PRD}{21}{2848}{1980}.

\bibitem{Hosotani1983}
Y.\ Hosotani, 
{\it ``Dynamical mass generation by compact extra dimensions''}, 
\jnl{\PLB}{126}{309}{1983};
{\it ``Dynamics of nonintegrable phases and gauge symmetry breaking''}, 
\jnl{\AP}{190}{233}{1989}.


\bibitem{Davies1988}
A.~T.~Davies and A.~McLachlan,
{\it ``Gauge group breaking by Wilson loops''},
\jnl{\PLB}{200}{305}{1988};
{\it ``Congruency class effects in the Hosotani model''},
\jnl{\NPB}{317}{237}{1989}.




\bibitem{Hatanaka1998}
H.\ Hatanaka, T.\ Inami and C.S.\ Lim,
{\it ``The gauge hierarchy problem and higher dimensional gauge theories''}, 
\jnl{\MPLA}{13}{2601}{1998}.




\bibitem{FiniteT2021}
 S.~Funatsu, H.~Hatanaka, Y.~Hosotani, Y.~Orikasa and N.~Yamatsu,
{\it ``Electroweak and left-right phase transitions in  $SO(5)\times U(1) \times SU(3)$ gauge-Higgs unification''}, 
\jnl{\PRD}{104}{115018}{2021}.

\bibitem{AnomalyFlow1}
 S.~Funatsu, H.~Hatanaka, Y.~Hosotani, Y.~Orikasa and N.~Yamatsu,
{\it ``Anomaly flow by an Aharonov-Bohm phase''}, 
\jnl{\PTEP}{2022}{043B04}{2022}, arXiv:2202.01393 [hep-ph].

\bibitem{AnomalyFlow2}
Y.~Hosotani,
{\it ``Universality in anomaly flow''}, 
\jnl{\PTEP}{2022}{073B01}{2022}, arXiv:2205.00154 [hep-th].


\bibitem{RS1999}
L.~Randall and R.~Sundrum, 
{\it ``A large mass hierarchy from a small extra dimension''}, 
\jnl{\PRL}{83}{3370}{1999}.


\bibitem{GUTinspired2019}
 S.~Funatsu, H.~Hatanaka, Y.~Hosotani, Y.~Orikasa and N.~Yamatsu,
{\it ``GUT inspired  $SO(5)\times U(1) \times SU(3)$ gauge-Higgs unification''}, 
\jnl{\PRD}{99}{095010}{2019}.


\bibitem{Bouchiat1972}
C.\ Bouchiat, J.\ Iliopoulos and Ph.\ Meyer,
{\it ``An anomaly-free version of Weinberg's model''}, 
\jnl{\PLB}{38}{519}{1972}.

\bibitem{GrossJackiw1972}
D.J.\ Gross and R.\ Jackiw,
{\it ``Effects of anomalies on quasi-renormalizable theories''}, 
\jnl{\PRD}{6}{477}{1972}.


\bibitem{CouplingSumRule}
  Y.~Hosotani, S.~Funatsu, H.~Hatanaka, Y.~Orikasa and N.~Yamatsu,
{\it ``Coupling sum rules and oblique corrections in gauge-Higgs unification''}, 
arXiv:2303.16418 [hep-ph].


\bibitem{CallanHarvey1985}
C.G.\ Callan, Jr.\ and J.A.\ Harvey,
{\it ``Anomalies and fermion zero modes on strings and domail walls''},
\jnl{\NPB}{250}{427}{1985}.

\bibitem{Fukaya2017}
H.\ Fukaya, T.\ Onogi and S.\ Yamaguchi,
{\it ``Atiyah-Patodi-Singer index from the domain-wall fermion Dirac operator''}, 
\jnl{\PRD}{96}{125004}{2017}.

\bibitem{WittenYonekura2021}
E.\ Witten and K.\ Yonekura,
{\it ``Anomaly inflow and the $\eta$-invariant''}, 
arXiv:1909.08775 [hep-th].






\end{thebibliography}
\end{document}